# Science and Technology Advance through Surprise

Feng Shi, University of North Carolina at Chapel Hill, Knowledge Lab
fbillshi@gmail.com

James Evans, Knowledge Lab, University of Chicago, Santa Fe Institute
jevans@uchicago.edu

Breakthrough discoveries and inventions involve unexpected combinations of *contents* including problems, methods, and natural entities, and also diverse *contexts* such as journals, subfields, and conferences. Drawing on data from tens of millions of research papers, patents, and researchers, we construct models that predict next year's content and context combinations with an AUC of 95% based on embeddings constructed from high-dimensional stochastic block models, where the improbability of new combinations itself predicts up to 50% of the likelihood that they will gain outsized citations and major awards. Most of these breakthroughs occur when problems in one field are unexpectedly solved by researchers from a distant other. These findings demonstrate the critical role of surprise in advance, and enable evaluation of scientific institutions ranging from education and peer review to awards in supporting it.

19[th] Century philosopher and scientist Charles Sanders Peirce argued that neither the logics of deduction nor induction alone could characterize the reasoning behind path-breaking new hypotheses in science, but rather their collision through a process he termed abduction. Abduction begins as expectations born of theory or tradition become disrupted by unexpected observations or findings (*1*). Surprise stimulates scientists to forge new claims that make the surprising unsurprising. Here we empirically demonstrate across the life sciences, physical sciences and patented inventions that, following Peirce, surprising hypotheses, findings and insights are the best available predictor of outsized success. But neither Peirce nor anyone since has specified where the stuff of new hypotheses came from. One account is serendipity or making the most of surprising encounters (*2, 3*), encapsulated in Pasteur's oft-quoted maxim "chance favors only the prepared mind" (*4*), but this poses a paradox. The successful scientific mind must simultaneously know enough within a scientific or technological context to be surprised, and enough outside to imagine why it should not be surprised. Here we show how surprising successes systematically emerge across, rather than within researchers; most commonly when those in one field surprisingly publish problem-solving results to audiences in a distant other. This contrasts with research that focuses on inter- and multi-disciplinarity as primary sources of advance (*5–7*). We suggest how predictability and surprise in science and technology provide us with new tools to evaluate how scientific institutions ranging from awards and graduate education to peer review facilitate advance.

In order to identify the sources of scientific and technological surprise, we must first identify what is expected with precision. Here we follow others in modeling discovery and invention as combinatorial processes linking previous ideas, phenomena and technologies (*8–12*). We separate combinations of scientific contents and contexts in order to refine our expectations about normal scientific and technological developments in the future (*13*). A new scientific or technological configuration of contents—phenomena, concepts, and methods—may surprise because it has never succeeded before, despite having been considered and attempted. A new configuration of contents that cuts across divergent contexts—journals and conferences—may surprise because it has never been imagined. The separate consideration of contents and contexts allows us to contrast scientific discovery with technological search: Fields and their boundaries are clear and ever-present for scientists at all phases of scientific production, publishing and promotion, but largely invisible for technological invention and its certification in legally protected patents.

Virtually all empirical research examining combinatorial discovery and invention has deconstructed new products into collections of pairwise combinations (*11*), resting on mature analysis tools for simple graphs that define links between entity pairs. Recent research, however, has demonstrated the critical importance of higher-order structure in understanding complex networks, ranging from the hub structure of global transportation networks to clustering in neuronal networks (*14*) to stabilizing interaction between species (*15, 16*). Here we develop a method to model the frontiers of science and technology as a complex hypergraph drawn from an embedding of contents and contexts (*17*) using mixed-membership, high-dimensional stochastic block models, where each discovery or invention can be rendered as a complete set of scientific contents and contexts. We demonstrate that adding this higher-order structure both improves our prediction of new articles and patents and those that achieve outsized success.

In this study, we apply our framework to three major corpora of scientific knowledge and technological advance: 19,916,562 biomedical articles published between 1865 - 2009 from the MEDLINE database; 541,448 articles published between 1893 - 2013 in the physical sciences from journals published by the American Physical Society (APS), and 6,488,262 patents granted between 1979 - 2017 from the US Patent database. The building blocks of content for those articles and patents are identified using community-curated ontologies—Medical Subject Heading (MeSH) terms for MEDLINE, Physics and Astronomy Classification Scheme (PACS) codes for APS, and United States Patent Classification (USPC) codes for patents (see SM Materials and Methods for details). Then we build a hypergraph for each dataset in each year where each node represents a code from the ontologies and each hyperedge corresponds to a paper or patent that inscribes a combination of those nodes, and compute node embeddings in those hypergraphs.

We build corresponding embeddings and hypergraphs of context where nodes represent journals, conferences, and major technological areas (for patents) that scientists and inventors draw upon in generating new work. Each hyperedge corresponds to a paper or patent that inscribes a combination of context nodes cited in its references. To predict new combinations, we develop a generative model that extends the mixed-membership stochastic block model (*30*) into high-dimensions, probabilistically characterizing common patterns of complete combinations (Figure 1). We model the likelihood that contents or contexts become combined as a function of their (1) *complementarity* in a latent scientific space and (2) cognitive *availability* to scientists through prior usage frequency. Specifically, each node $i$ is associated with a latent vector $\theta_i$ that embeds the node in a latent space constructed to optimize the likelihood of the observed papers and patents. Each entry $\theta_{id}$ of the latent vector denotes the probability that node $i$ belongs to a latent dimension $d$. (The latent vectors capture but do not replicate the ontology structures; see Fig. S1.) The complementarity between contents or contexts in a combination $h$ is modeled as the probability that those nodes load on the same dimensions, $\sum_d \prod_{i \in h} \theta_{id}$. We also account for the cognitive availability of each content and context as most empirical networks display great heterogeneity in node connectivity, with a few popular contents and contexts intensively drawn upon by many papers and patents. Accordingly, we associate each node $i$ with a latent scalar $r_i$ to account for its cognitive availability or the exposure scientists have had to it, measuring its overall connectivity in the network. The propensity ($\lambda_h$) of combination $h$—our expectation of its appearance in actual papers and patents—is then modeled as the product of the complementarity between the nodes in $h$ and their availability:

$$\lambda_h = \sum_d \prod_{i \in h} \theta_{id} \times \prod_{i \in h} r_i \ .$$

Then the number of publications or patents that realize combination $h$ is modeled as a Poisson random variable with $\lambda_h$ as its mean. Finally, the likelihood of a hypergraph $G$ is the product of the likelihood of observing every possible combination (see SM Materials and Methods for details).

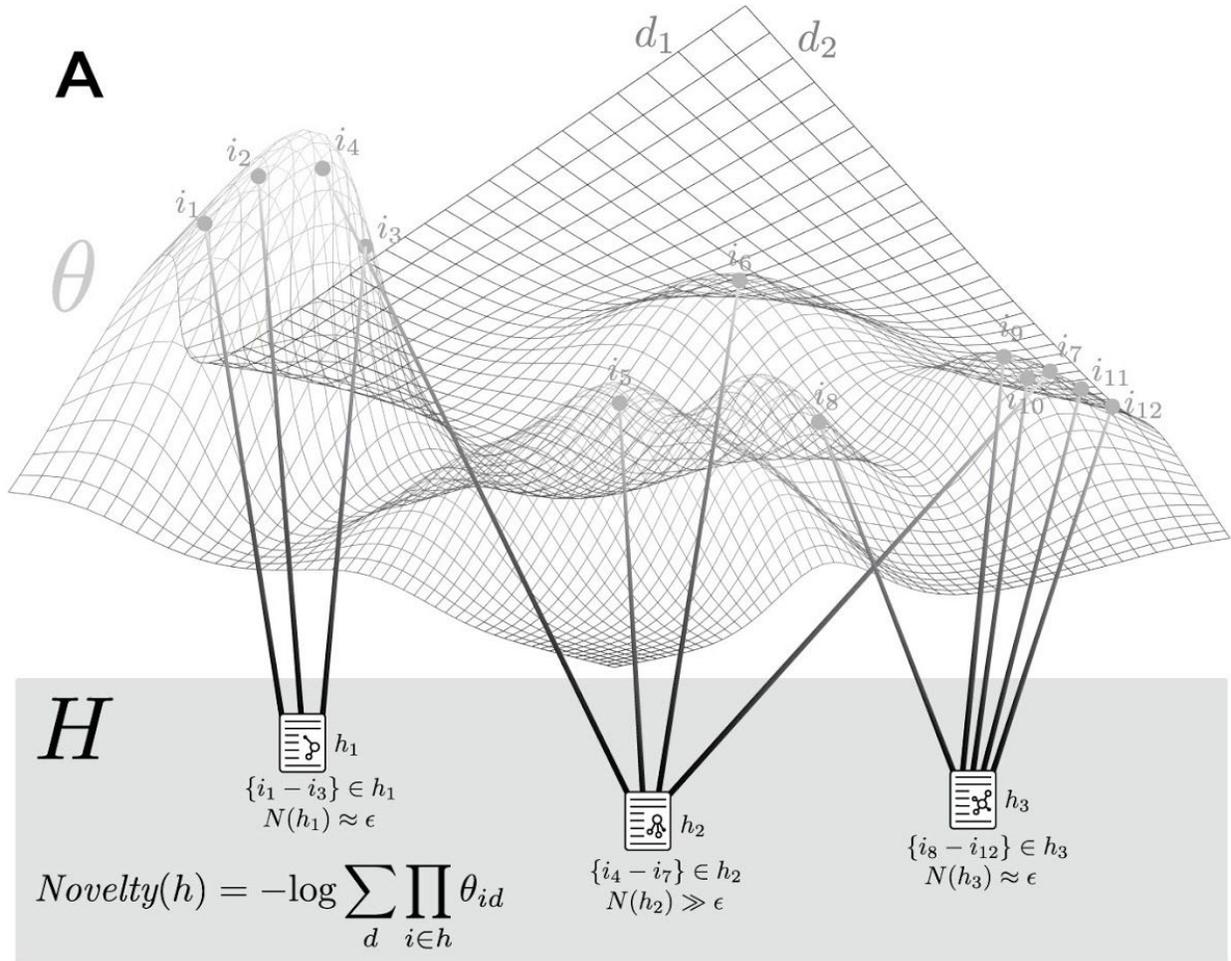

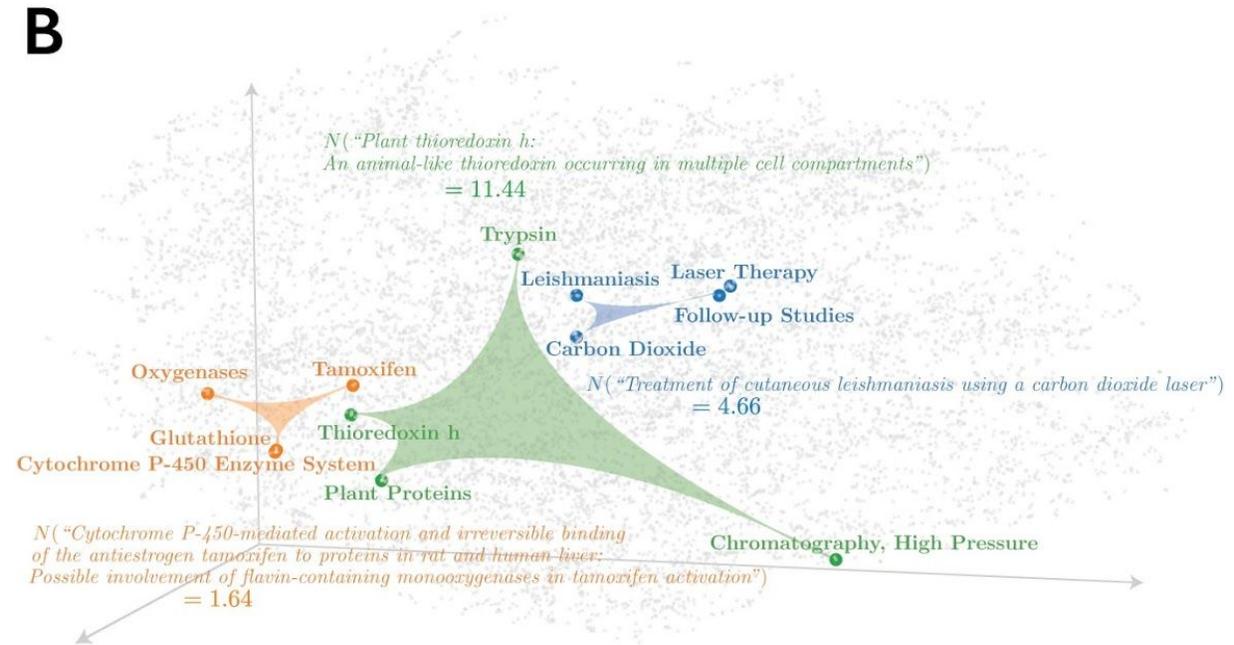

**Figure 1. (A)** Illustration of the manifold inscribing all embeddings θ and an evaluation of three articles or patents (hyperedges

$h_{1-3}$) in terms of their novel combinations. Articles/patents $h_1$ and $h_3$ represent projects that combine scientific or technical components near one another in θ, making each of high probability and low (ε) novelty—similar to many related papers from the past. By contrast, paper $h_2$ draws a novel combination of components unlike any paper from the past, making it of low probability and high ($\gg$ ε) novelty. See Fig. S2 for a real density plot of the MeSH terms. **(B)** Actual three dimensional projection of the embeddings of a sample of MeSH codes from MEDLINE articles in our analysis. Also included are MeSH terms in the most novel article (blue), the least novel article (orange), and a random article in between (green) among this sample of articles including four MeSH terms.

Across biomedical sciences, physical sciences, and inventions, the model correctly distinguishes between a content combination that turned into a publication and a random combination more than 95% of the time based on data from previous years (Biomedicine: AUC=0.98; Physics: AUC=0.97; Inventions: AUC=0.95; see SM Materials and Methods for model evaluation). New context combinations are also predictable (Biomedicine: AUC=0.99; Physics: AUC=0.88; Inventions: AUC=0.83). The model implies that researchers tend to conservatively wander locally across the latent knowledge space constructed from papers and patents in prior years to arrive at those published the following year. This agrees with previous findings on inertia in scientific and technological investigations as teams wander locally across contents and contexts to extend their own and colleagues' prior work (*18*).

With a measure of what science and technology is common and expected, we assess surprising, improbable combinations as the surprisal of $h$ (*19*, section 3.3):

$$novelty(h) = -\log \sum_d \prod_{i \in h} \theta_{id}. \tag{2}$$

We first examine how novelty is associated with citation impact and awards by dividing the MEDLINE papers into 10 equal-sized groups in ascending order of citation count. We also normalize novelty scores by transforming them into percentiles. The group-average content and context novelties both increase significantly with citation decile, as shown in Figure 2 (A and B). The same pattern is observed in APS papers and patents (Fig. S3 and S4). Further, we show that Nobel prize-winning papers, which are all in the top 10% citation group, have lower-than-average context novelty, but extremely high content novelty. For general awards in Biology and Medicine, most award-winning papers in the top 10% group follow this pattern. This divergence between citations and awards for papers with high context novelty is likely because citations are conferred by everyone who credits an advance, but awards are offered by particular scientific communities or contexts, which apparently undervalue breakthrough advances that transgress established boundaries (*20*).

Moreover, the probability of being a hit paper—in the top 10% of most cited papers published in the same year—also increases monotonically with the percentile of novelty. For MEDLINE papers, those with the most novel combinations of context are on average 4 times more likely to be a hit paper than random, novel content combinations are 2 times more likely (Figure 2C), and the most novel of content and context jointly are approximately 5 times more likely (Figure 2D). MEDLINE papers with maximal joint context and content surprise predict nearly 50% of the likelihood of being in the top 10% of citations.

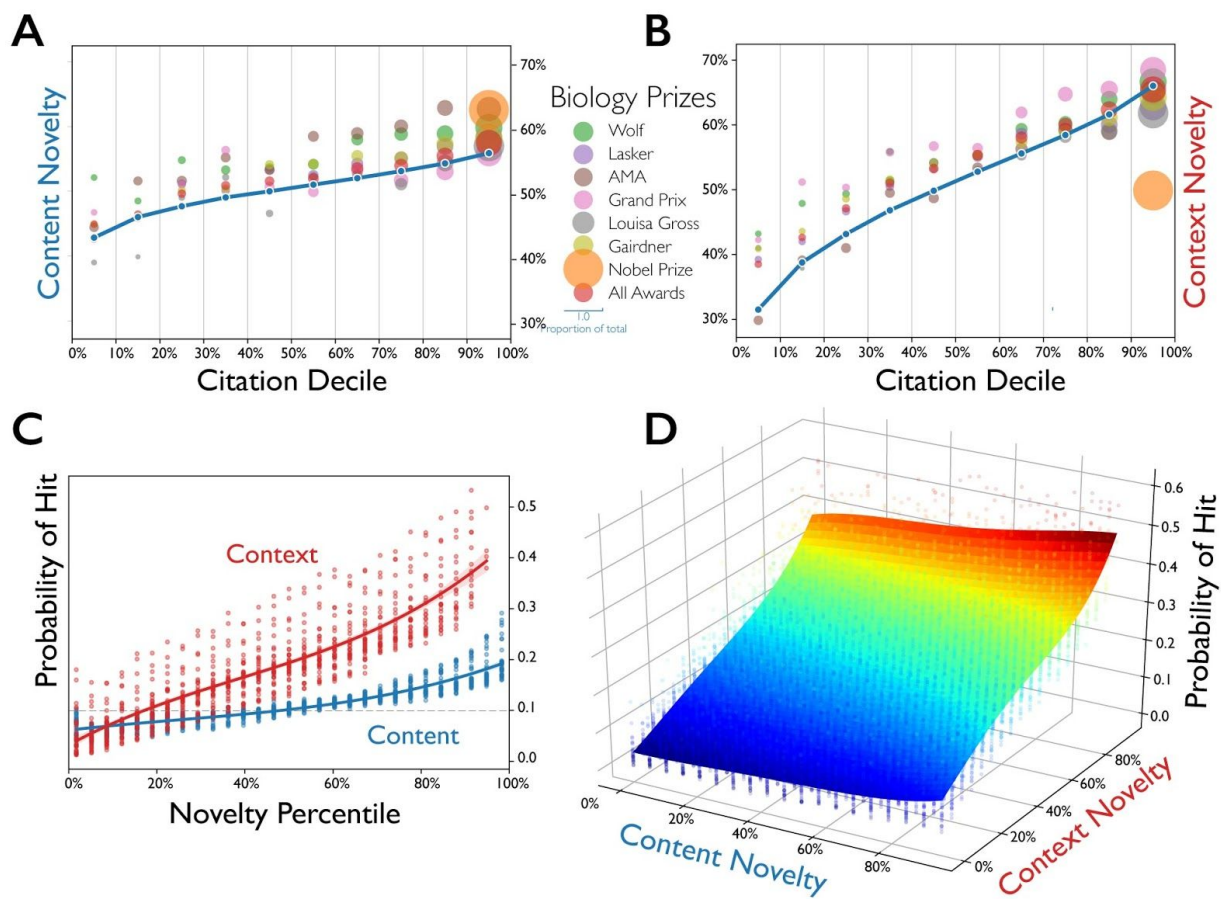

**Figure 2**: Association between novelty and citation impact or awards for MEDLINE papers. Average content (A) and context (B) novelty are plotted for each decile of citations, tracing a monotonic rise; Including averages for Nobel prizes in Physiology or Medicine and general awards in Biology and Medicine. Probability of being a hit paper is plotted against content and context novelty separately (C) and jointly (D), manifesting a monotonic increase with novelty.

Unlike MEDLINE papers, novel patents are only 2 times more likely to be a hit patent than random (Fig. S4). Disciplinary boundaries are weaker in the technology space, where patent examiners, unlike scientific reviewers, do not enforce them. The lack of discrete fields enables technologists to search more widely, but reduces the signal from violations of context in the prediction of advance.

Both content and context novelties are good predictors of impact, but they provide nearly independent information regarding the ongoing construction of scientific ideas and technological artifacts. The correlation between propensities for content and context combinations are extremely low across MEDLINE (0.01), APS (0.05), and patents (0.03). When we calculate the content similarity between cited journals and publishing venues where the citing papers are published (see SM Supplementary Text for details), we see that scientists cite contexts similar to publishing venues 500% more intensively than contexts that are distant (Figure 3 Left) (*23*). Inventors of patented technologies, however, are not reviewed by peers and cite close or distant sources with roughly the same probability. Following from this difference, we find that the distribution of collective attention differs dramatically in science versus technology. We quantify the spread of attention with the normalized entropy of the number of publications containing each content node, shown in Figure 3 Right. Content nodes in the patent space receive more equal attention (higher entropy), compared to MEDLINE or APS.

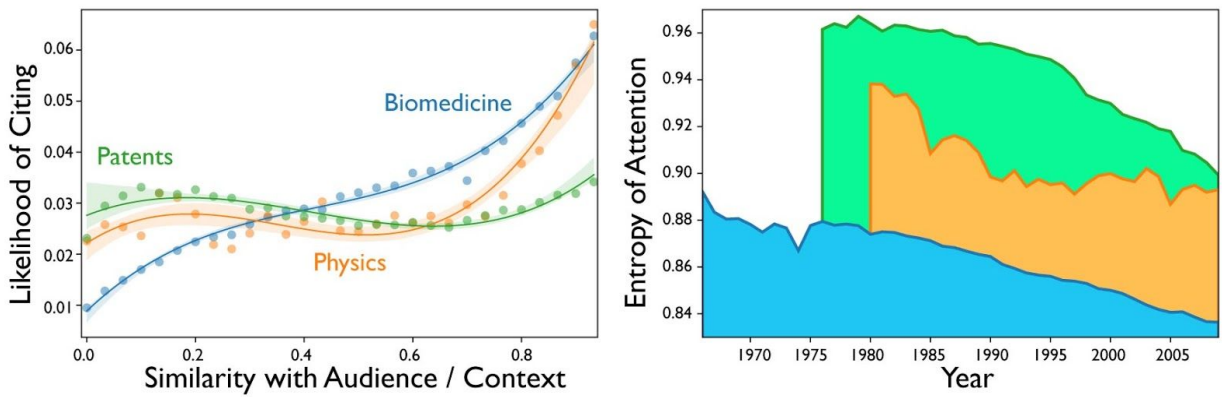

**Figure 3:** Left: The likelihood of citing context nodes variously familiar within the publication venue for papers in MEDLINE (blue), APS (orange), and US Patent (green). Papers in MEDLINE and APS reference contexts similar to those in which they are published much more intensively than contexts that are distant. Patents, by contrast, reference close or distant sources with roughly the same likelihood. Right: Entropy of attention on the content nodes over time. The entropy of attention is calculated as the entropy of the number of publications associated with each content node. To compare entropy across datasets, it is normalized by the logarithm of the number of content nodes in each dataset. The content nodes in the patent space receive more equal attention (higher entropy), compared to MEDLINE and APS, across the years shown in the figure.

Finally, we explore the relationship between scientists' backgrounds and breakthrough. Do unusual individual scientist backgrounds, atypical collaborations, or unexpected expeditions where scientists and inventors reach across disciplines and address problems framed by a distant audience contribute most to novelty and impact? Using context (e.g., journals, conferences) embeddings, $\theta_i$, and Eq. 2, we quantify (1) *career novelty* of a scientist by the surprisal of the combination of contexts she has ever published, (2) *team novelty* by the combination of contexts brought together across team members' publication histories, and (3) *expedition novelty* by the average distance between the backgrounds of team members and their audience formalized by their publication venue. Figure 4A shows that, for MEDLINE papers the probability of being a hit paper increases gradually with career and team novelty, but expedition novelty rises much more quickly as the strongest predictor. Papers involving the most unexpected publication events are 3.5 times more likely than random to be hit papers. Three-dimensional novelty distributions graphed in Figure 4B also show that career and team novelties are highly correlated, suggesting that successful teams not only have members from multiple disciplines, but also members with diverse backgrounds who stitch interdisciplinary teams together. Successful knowledge expeditions, however, are the most likely path associated with breakthrough discovery. Figure 4B also reveals how high expedition novelty in the absence of team and career novelty still remains associated with an increased probability of hit papers. The pattern of effects in APS is consistent with that in MEDLINE where the association between expedition novelty and the probability of hit papers is even more pronounced, relative to team and career novelty (Fig. S5). Unsurprisingly, the pattern is different for patents (Fig. S6) where expedition novelty is not significantly associated with hits because novelty is the primary basis of evaluation, subfields are not enforced and, as a result, expedition novelty is so frequent that it loses its value as a signal of outsized success: its skewness is .61, nearly three times larger than in the life sciences (skewness 0.26) and the flipside of physics, where such expeditions are uncommon (-0.36).

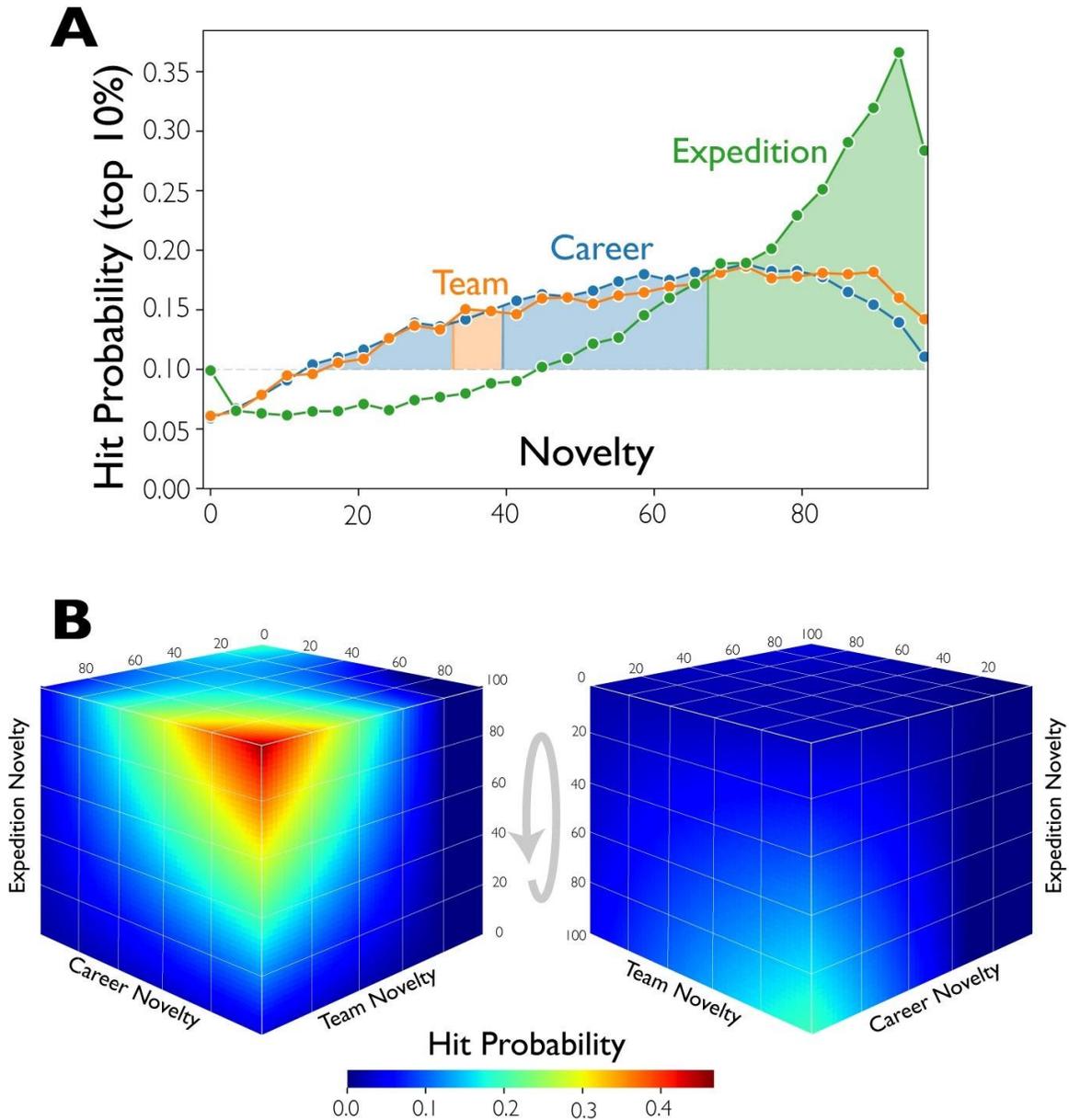

**Figure 4:** Association between scientists' backgrounds and impact. A: The probability that a hit biomedical paper was produced by scientists manifesting greater career, team and expedition novelty; with career and team novelty closely correlated and expedition novelty sharply deviating. B: Hit probability as a function of career, team, and expedition novelties jointly with hit probability denoted by color.

In this paper we demonstrate the striking predictability of future scientific articles and technology patents, which results from a system in which researchers, their collaborators, students, and fields produce self-similar streams of research over time. We then justify the importance of surprise in unfolding discovery and invention, revealing that up to 50% of outsized success (in biomedicine) can be predicted by improbability under models that predict new research products. Most of those unpredictable successes occurred not necessarily through interdisciplinary careers or multi-disciplinary teams, but from scientists in one domain solving problems in a distant other. This implies the operation of collective abduction, where violations of theoretical and traditional expectations drive collective attention. It further suggests the cross-disciplinary search process by which problems, puzzles and conflicts in one area of science become discovered by scientists in other areas whose exposure to foreign theories and findings enable them to make surprising discoveries.

We also technically demonstrate that prediction of complex content and context bundles dramatically benefit from taking into account the high-dimensional structure of complete combinations, rather than viewing them as sets of pairwise combinations. This suggests the potential importance of representing high-dimensional structure like sets in a form that captures their native complexity for characterization and prediction.

Our findings suggest how models that predict normal and outsized advance represent powerful tools for evaluating the degree to which scientific and technical institutions facilitate progress. For example, our work shows that granting scientific awards for breakthrough progress, from Nobel Prizes to the plaques and certificates sponsored by nearly every scientific society are biased towards some forms of surprise and away from others. Scientific societies convene conferences and publish journals, the central contexts that showcase new findings, and so it is notable that they tend to award surprising combinations of scientific contents and not contexts, but that novel context combinations are more predictive of outsized citations and possibly scientific importance. This suggests that awards, as currently offered, represent a conservative influence on scientific advance (*20*). Similarly, our findings reveal that scientists amplify the familiarity of their work to colleagues, editors and reviewers, increasing their references to familiar sources by nearly 500%, likely in order to appear to build on the shoulders of their audience. This reinforces the internal focus of dense fields, which collectively learn more about less. Inventors, by contrast, cite and search widely to know less about more (*24*), providing new evidence for complementarities between search in science and technology and justifying why fundamental insights emerge not only from fundamental investigations, but also practical ones (*25–27*). Finally, our work has implications for graduate education. Novel careers, novel combinations of experience within teams, but most critically, researchers seeking out problems and subjects held by distant audiences, if successful, dramatically increase the likelihood that their work will disrupt scientific attention with insights received as path breaking. This suggests that education seeking to cultivate scientific breakthroughs might teach trans-disciplinary search for problems, and frame every student, team and expedition as an experiment, whose complex combination of background experiences could condition novel hypotheses with the potential not only to succeed or fail, but to radically alter science.

## Supplementary Materials

**Materials and Methods**

Data Overview

This work investigated three major corpora of scientific and technological knowledge: 19,916,562 papers published between 1865 - 2009 in biomedical sciences from the MEDLINE database, 541,448 papers published between 1893 - 2013 in physical sciences from all journals published by the American Physical Society, and 6,488,262 patents granted between 1979 - 2017 from the US Patent database (USPTO). The building blocks of content for those articles and patents are identified using community-curated ontologies—Medical Subject Heading (MeSH) terms for MEDLINE, the Physics and Astronomy Classification Scheme (PACS) codes for APS, and United States Patent Classification (USPC) codes for patents. Then we build hypergraphs of content where each node represents a code from the ontologies and each hyperedge corresponds to a paper or patent that realizes a combination of the nodes.

We acknowledge the potential conservative influence from using established keyword ontologies rather than all of the words from titles, abstracts or full-text of articles and patents. Nevertheless, we note that the ontologies we examine do evolve over time, with active additions following the concentration of research in a given area. Moreover, these ontologies allow us to use the community of authors (APS), annotators (MEDLINE), and examiners (USPTO) to crowdsource the disambiguation of scientific and technological terms. Future work may explore how words differ from keywords, especially in the emergence of new fields.

MEDLINE

MEDLINE is the U.S. National Library of Medicine's bibliographic database. It contains abstracts, citations, and other metadata for more than 25 million journal articles in biomedicine and health, broadly defined to encompass those areas of the life sciences, behavioral sciences, chemical sciences, and bioengineering. The version of data used in this study contains 19,916,562 papers published between 1865 - 2009. Because the coverage for papers prior to 1966 is somewhat limited, our analysis focuses on papers published in and after 1966, but with the pre-1966 papers as background information when predicting new content and context combinations and their novelty.

Medical Subject Headings (MeSH) is the National Library of Medicine's (NLM's) controlled terminology used for indexing articles in MEDLINE. It is designed to facilitate the determination of subject content in the biomedical literature. MeSH terms are organized hierarchically as a tree with the top level terms (called headings) corresponding to major branches such as "Diseases" and "Chemicals and Drugs", with multiple levels under each branch. Terms in the bottom level are the most fine-grained, detailed concepts associated with distinct biological phenomena, chemicals, and methods. We use the bottom-level terms from the 3 branches that are central to the biomedical field - "Diseases", "Chemicals and Drugs", and "Analytical, Diagnostic and Therapeutic Techniques and Equipment" - as nodes in the hypergraphs of content of MEDLINE papers. Terms from the Diseases branch include conditions such as "lathyrism" and

"endometriosis"; examples from the Chemicals and Drugs branch include "elastin", "tropoelastin", "aminocaproates", "aminocaproic acids", "amino acids", "aminoacetonitrile", and "amyloid beta-protein"; and examples from the Analytical, Diagnostic and Therapeutic Techniques and Equipment branch (or methods for short) include "polyacrylamide gel electrophoresis", "ion exchange chromatography", and "ultracentrifugation". NLM curators manually affix MeSH codes to papers as they are ingested into MEDLINE and made available through the popular PubMed platform.

APS

The APS dataset is released by the American Physical Society (APS). It contains 541,448 papers published between 1893 and 2013 in 12 physics journals: *Physical Review*, *Physical Review A, B, C, D, E, I* and *X*, *Physical Review Special Topics - Acceler and Physics*, *Physical Review Letters*, and *Reviews of Modern Physics*.

The dataset contains basic metadata for each paper including title, publication year, abstract, etc. It also contains the PACS (Physics and Astronomy Classification Scheme) codes associated with each paper. We use the PACS codes as nodes in hypergraphs of content to characterize APS papers. The Physics and Astronomy Classification Scheme was developed by the American Institute of Physics in 1970 and has been used by APS since 1975. Similar to MeSH terms, PACS is also a hierarchical partition of the whole spectrum of subject matter in physics, astronomy, and related sciences. Since PACS codes are not available for papers published before 1975, our analysis is restricted to APS papers published in and after 1975. Like MeSH terms, PACS codes are arranged hierarchically, and include "Mathematical methods in physics", which range from "Quantum Monte Carlo Methods" to "Fourier analysis"; "Instruments…" such as "Electron and ion spectrometers" and "X-ray microscopes"; "Specific theories…" like "Quark-gluon plasma" and "Chiral Langrangians"; and "...specific particles" ranging from "Baryons" to "Quarks". Unlike MeSH codes, which are added by curators, authors affix PACS codes to their own papers through the publishing process.

The dataset only contains citations between the APS papers. In order to obtain external citations we query the Web of Science (WOS) database to collect all the journals cited by the APS papers. Particularly, in the WOS database we find all the papers published by the 12 APS journals, and then all the journals cited by those papers. The journals are then used as nodes in hypergraphs of context for the APS papers. Additionally, we also query the WOS database to collect the number of citations a paper receives for more accurate assessment of the paper's impact.

US Patent

The US Patent dataset is released by the US Patent & Trademark Office (USPTO). It contains 6,488,262 patents published between 1979 and 2017. The dataset contains basic metadata for each patent such as title, publication year, USPC (United States Patent Classification) codes, etc. The USPC is a classification system used by USPTO to organize all U.S. patent documents and other technical documents into specific technology groupings based on common subject matter. The USPC is a two-layer classification system. The top layer consists of terms called classes, and each class contains subcomponents called subclasses.

According to USPTO, a class generally delineates one technology from another and every patent is assigned a main class. As such, we use the class codes as nodes in the hypergraphs of context for patents. Subclasses delineate processes, structural features, and functional features of the subject matter encompassed within the scope of a class, and thus we use subclass codes as content nodes for the patents. In total, there are 158,073 subclass codes (content nodes) and 496 class codes (context nodes).

On January 1, 2013, the USPTO moved to a new classification system called the Cooperative Patent Classification (CPC); consequently, our analysis is restricted to patents granted before 2013.

Nobel Prize Papers

The Nobel prize-winning papers are derived from the Nobel laureates dataset by Li et al. (29), which contains publication histories of nearly all Nobel prize winners from the past century. However, their focus is on the Nobel laureates, but ours is on award-winning papers. While it is relatively easy to find out the person who won a prize, it is hard to pinpoint the papers that contribute to the winning of the prize. Li et al. take a generous approach by including papers cited by Nobel lectures and papers published in the same period of one's prize-winning work (while satisfying several inclusion criteria; see (29) for details). This results in noises for our analysis as not every paper in their dataset is a prize-winning paper. As a conservative solution, for every Nobel laureate we take the most cited paper in the dataset as the award-winning paper and use only those papers as award-winning papers in our analysis. We acknowledge that a few Nobel prizes are attributed to an opus of work and this filtering process might miss a few relevant papers, but the most important (in terms of impact) paper for every prize is kept and every paper remaining represents an award-winning paper.

General Award-Winning Papers

In Fig. 2 and S3, we compare the content and context novelty of our entire population of research papers with those associated with awards. The general awards data is from Foster et al. (32). They define award-winning papers as those authored by a scientist who won an international award or prize within 30 years. They first identified the winners of 137 different prizes and awards from biology, medicine, and chemistry. To create a large list of prizes, they drew from the category pages for biology awards, medicine awards, and chemistry awards in Wikipedia and then validated them with several biomedical scientists:

en.wikipedia.org/wiki/Category:Biology_awards
en.wikipedia.org/wiki/Category:Medicine_awards
en.wikipedia.org/wiki/Category:Chemistry_awards.

Note that more and less prestigious prizes are gathered on these category pages, with the vast majority of prestigious prizes being captured (Nobel Prizes in Physiology or Medicine or Chemistry, a range of prizes considered "pre-Nobel", etc.). To eliminate prize winning papers that were not associated with high-achievement, they deleted those prizes whose selection criteria were not research related (e.g., prizes awarded for teaching) as well as prizes targeted at students. For each research-related prize, they searched for the names of prize winners in PubMed and if at all possible identified three articles written by each winner. Those awards that center on biomedical contributions have deep coverage in MEDLINE articles and so necessarily represent more data points in our analysis than awards peripheral to biomedicine (e.g., The Kempe Award for Distinguished Ecologists). With this expectation, they also devoted more time to thoroughly completing the coverage of articles associated with major or broad awards, which we believed should be well represented in MEDLINE. Hence, these major and broad awards are necessarily more complete (i.e., more prize winners have associated articles) than narrower awards or those that only partially overlap the biomedical focus of PubMed (e.g., entomology).

This collection of articles provided the seed for a broader search. They identified the name of the prize winner in the MEDLINE entry for each of the one to three seed articles. Using the Author-ity tool (33), this focal name can be mapped to a cluster of names with partial disambiguation. They searched MEDLINE for the names in each of the author clusters that map to the prize winner's name in the one to three hand-assigned articles. They then assigned all papers retrieved in this way to the prize winner, if they were written from 0 year to 30 years before the award was granted; see Table S2 in (18) for a list of prizes and number of associated papers discovered. If their research assistants misassigned any of the one to three seed articles, this process could introduce false positives for common names (the Author-ity author clustering is quite conservative and is unlikely to contribute to this error). As these false positives would be "typical" rather than award-winning papers, however, they likely behave like the majority and dilute the effect we report. Papers written by authors with non-English characters in their names are also underrepresented. These false negatives similarly dilute our estimate of the distinctiveness of award-winning papers. Our non-Nobel prize findings should therefore be viewed as a conservative estimate of the difference between prize winning papers and the pool of all papers. Note that prize winning papers are significantly higher in content novelty than the average paper within a given citation decile, but that they are not systematically higher in context novelty, likely because awards are typically conferred by a context—a journal, association, or field.

Higher-Order Stochastic Block Model

For a given hypergraph, whether comprised of content or context nodes, the propensity of any combination of nodes to form a hyperedge is modeled as a product of two factors: the complementarity between the nodes in the combination and their cognitive availabilities. Combinations with higher propensity will be more likely to turn into papers and patents, agreeing with the intuition that people tend to search locally and pursue trending topics.

To formulate this idea formally, each node $i$ is associated with a latent vector $\theta_i$ that positions the node in a latent space constructed to optimize the likelihood of observed papers and patents. Each entry $\theta_{id}$ of the latent vector denotes the probability that node $i$ belongs to a latent dimension $d$, and thus $\sum_{d=1}^{D} \theta_{id} = 1$. The complementarity between nodes in a combination $h$ is modeled as the probability that those nodes belong to the same dimension, $\sum_{d} \prod_{i \in h} \theta_{id}$. This formulation represents an extension of the mixed-membership stochastic block model in (*30*), which was designed for networks with only pairwise interactions.

We also account for each node's cognitive availability because most empirical networks display great heterogeneity in node connectivity, with few contents intensively drawn upon and few contexts widely attended or appreciated across many papers and patents. Previous work (*31*) has shown that by integrating heterogeneity of node connectivity, the performance of community detection in real-world networks dramatically improves. Accordingly, we associate each node $i$ with a latent scalar $r_i$ to account for its cognitive availability, presumably associated with its overall connectivity in the network.

Assembling these components, the propensity ($\lambda_h$) of combination $h$—our expectation of its appearance in actual papers and patents—is modeled as the product of the complementarity between the nodes in $h$ and their availabilities

$$\lambda_h = \sum_d \prod_{i \in h} \theta_{id} \times \prod_{i \in h} r_i. \tag{1}$$

To link the propensities to their observed appearances, we model the number of papers or patents $X_h$ that realize a certain combination $h$ as a Poisson random variable with the propensity of that combination as its mean:

$$X_h \sim Poisson(\lambda_h)$$

Accordingly, the probability of observing a hypergraph $G$ is the product of probabilities of observing all possible combinations:

$$P(G|\Theta, R) = \prod_{h \in H} P(x_h|\Theta, R),$$

where $x_h$ is the number of observed papers or patents that realize combination $h$ and $H$ is the set of all possible combinations. $(\Theta, R)$ denotes all unknown parameters: $\Theta = (\theta_1, ..., \theta_n)$ and $R = (r_1, ..., r_n)$.

Finally, we model a time sequence of hypergraphs $(G^1, ..., G^T)$ as the output of a Hidden Markov Process on latent parameters $\Theta, R$:

$$P(G^1, ..., G^T|\Theta^1, ..., \Theta^T, R^1, ..., R^T) = P(G^1|\Theta^1, R^1) \prod_{t=2}^{T} P(\Theta^t, R^t|\Theta^{t-1}, R^{t-1}) P(G^t|\Theta^t, R^t),$$

where time is indexed by the superscript $t$, and the transition density $P(\Theta^t, R^t|\Theta^{t-1}, R^{t-1})$ is a Gaussian density.

Given articles published by a certain year $T$, we estimate parameters $(\Theta^1, ..., \Theta^T, R^1, ..., R^T)$ by maximizing the likelihood function above via stochastic gradient descent. Then the model enables us to predict combinations in year $T + 1$. However, even with stochastic gradient descent, model estimation is still computationally challenging due to the vast space of possible combinations. We address these issues in the estimation process as follows. First, the space of possible combinations is exponentially large (on the order of $2^n$), and it is computationally prohibitive to go over all possible combinations even with stochastic gradient descent. However, it is extremely rare for large combinations to turn into hyperedges, and hence, we restrict the set of possible combinations to include only combinations no larger than the largest hyperedge observed. Second, since the real hypergraphs are sparse, the sets of hyperedges and non-hyperedge combinations are exceedingly unbalanced with the number of hyperedges to be on the order of $n$ but the number of non-hyperedge combinations on the order of $n^D$ (where $D$ is the size of the largest hyperedge). We employ a widely used approach, negative sampling, in machine learning to address this unbalance issue. Specifically, in each iteration of the training (optimization) process, we randomly sample as many non-hyperedge combinations as the hyperedges to construct balanced hyperedge and non-hyperedge sets. Lastly, to facilitate the stochastic gradient descent, we take a mini-batch of hyperedges and non-hyperedges to compute the gradient of the objective function at each step of the training process.

Model Evaluation

As a brief summary, we study 3 datasets: MEDLINE, APS, and US Patent; each dataset contains hypergraph data over several decades; and we model content and context hypergraphs separately. Consequently, we estimate hundreds of models with each model fitted to a specific type of hypergraph (content or context) from one of the three datasets up to a certain year. Then, we evaluate the fitness of each model by its predictive performance of (out-of-sample) future combinations.

For example, given hypergraphs of MeSH terms up to and including year 2008, we estimate the stochastic block model, and use the estimated model to predict hyperedges in 2009. Specifically, using the estimates of the parameters $(\theta, r)$ for year 2008, we can compute the propensity $\lambda_h$ of any combination $h$ of MeSH terms in year 2009, following Equation (1) $\lambda_h = \sum_c \prod_{i \in h} \theta_{ic} \times \prod_{i \in h} r_i$. Then we assess the model's predictive performance in terms of its AUC (Area Under the Operator-Receiver Curve). Statistically, AUC is the probability that a random combination which turned into a hyperedge (positive combination) in 2009 have a larger propensity than a random combination that did not turn into a hyperedge (negative combination) in 2009. To estimate this quantity, we randomly sample a positive combination and a negative combination of the same size from 2009, and check whether the positive combination has a larger propensity than the negative. The simulation is repeated for 10000 times and we calculate the fraction of times where the positive has larger propensity than the negative, which is the estimation of the AUC score in predicting hyperedges in 2009. It is easy to see that a perfect predictor would achieve an AUC score of 1 and random guesses would have an AUC of around 0.5. The larger the number, the better the predictive performance.

**Supplementary Text**

Probability of hit papers and patents

A hit paper (or patent) is defined as one among the top 10% most cited papers (or patents) published in the same year. For example, for all the papers published in 1990, we count all the citations they received in the time period covered by our dataset, and the top 10% most cited papers are hit papers.

To study the effect of novelty on the probability of being a hit paper, for all the papers published in each year, we first transform the raw novelty scores of the papers into percentiles, and then divide them into 30 equally sized bins in ascending order of the novelty scores. Then we assess the probability of hit papers for each bin as the fraction of hit papers in that bin. Finally, the probability of hit papers for each bin is plotted against the center percentile value in that bin in Figures 2, 4, S3, S4, S5 and S6. Similarly, to study the joint effect of different novelties on the probability of being a hit paper, we divide the papers into multiple bins according to the different novelties simultaneously. For example, for the joint effect of content and context novelties, we divide the paper into 30-by-30 bins in terms of their content and context novelties at the same time, and then calculate the fraction of hit papers in each bin.

Preference on context citations

To assess the extent to which scientists and inventors cite contexts (e.g., journals and conferences) that are familiar to their audience, we compute the similarity between every pair of context nodes where one cites the other. For example, for a paper $i$ published in journal $X$, we calculate the similarity between the journal $X$ and every journal cited by paper $i$. The similarity is quantified by the cosine similarity between two vectors representing the content of the two journals, conditioned on the content of paper $i$. Specifically, each journal is represented by a vector and each entry in the vector corresponds to a content node (MeSH terms, PACS codes, or subclasses); the value of an entry is the number of papers containing the corresponding content node and ever published by the journal, appropriately normalized so that the sum of the vector is 1. In other words, the vector consists of the loadings of the journal on different contents. When calculating the similarity between two journals, we don't directly compute the cosine similarity between their vectors, as the vectors contain a lot of information irrelevant to the paper currently under consideration. Instead, we only use the entries corresponding to the content nodes in paper $i$ to calculate the cosine similarity between the two journals.

As we sweep through all the papers (or patents), a distribution of the similarity between citing-citee context pairs is obtained: the number of times for which context nodes at a given similarity with the audience context (i.e., the citing context) are cited. To appropriately normalize this distribution, we also compute the potential space of citation similarity, which is the number of times for which context nodes at a given similarity would be cited at random. This is achieved by the following procedure: for each paper, sample as many context nodes uniformly at random from all the context nodes as those originally cited, treat the sampled context nodes as if they were cited by the paper, and carry out the same similarity calculations as above. Finally, we have two distributions of similarity between citing-citee context pairs - one observed and one simulated by random sampling - and we take the ratio of the two as the likelihood of citing a context at a given similarity with the audience's context.

Limitations

Our study provides recommendations for improving search in science and technology, but not designs for a machine that generates surprising future discoveries and inventions, because our model only predicts how surprising combinations that *succeed* at publication become success. A vast range of content, context, and background combinations that are nonsensical or doomed, if systematically pursued in a self-conscious search for surprise, would dramatically decrease the rate of outsized success we demonstrate here. Our high-dimensional treasure map does not show where X marks the spot, but rather powerfully reveals the regions that have been over-explored, where the likelihood of making a new discovery or invention is vanishingly low. In this way, our characterization of the high-dimensional space of discoveries and inventions, combined with the validation of surprise as a core principle of unfolding scientific and technology growth, reveals the possibility of negative crowdsourcing, where researchers can exploit the crowd estimate of prior fruitfulness to identify where not to look for important opportunities.

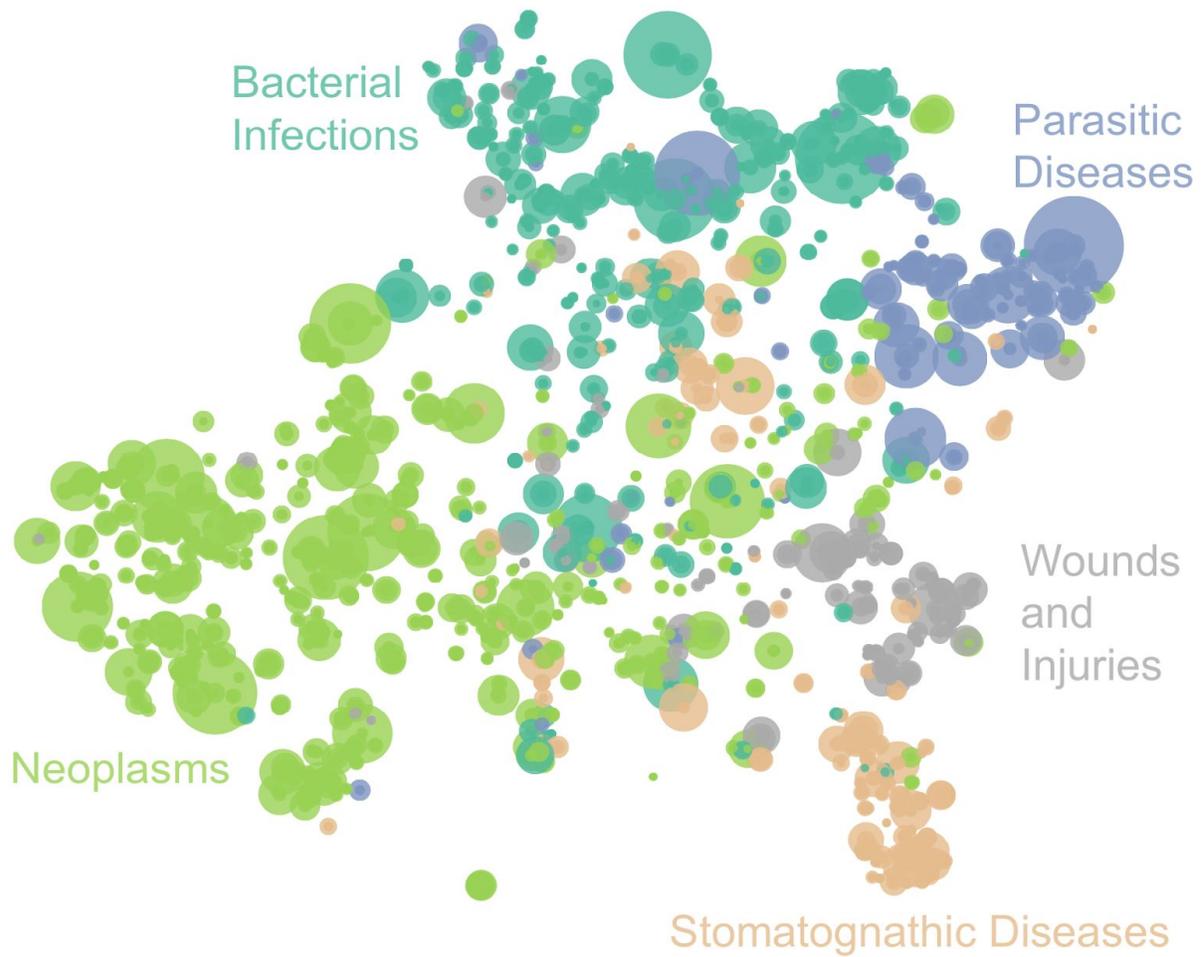

**Fig. S1.** 2D projection of the embeddings of diseases in a sample of MEDLINE papers published in 1990 using t-SNE. Each circle denotes a MeSH term for diseases, whose size is proportional to the number of papers associated with it and whose color corresponds to a top-level disease type.

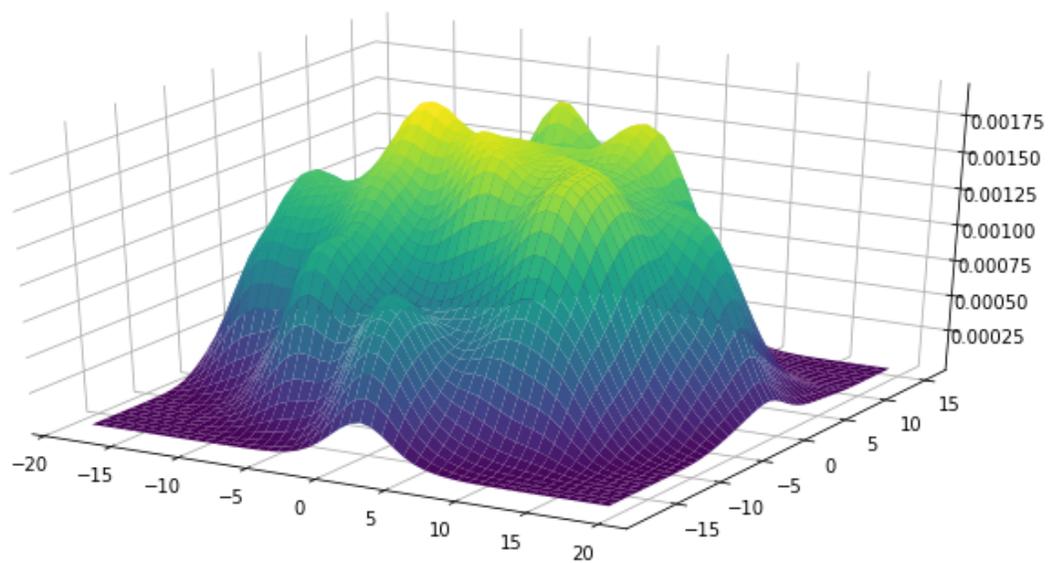

**Fig. S2.** Density plot of 2D projections of the MeSH term embeddings in papers published in 1990. We take all the MeSH terms that are active in 1990 (associated with any paper published in 1990) and project their high-dimensional embeddings onto a 2D plane using t-SNE; a Gaussian kernel density is then fit to the 2D points of the MeSH terms.

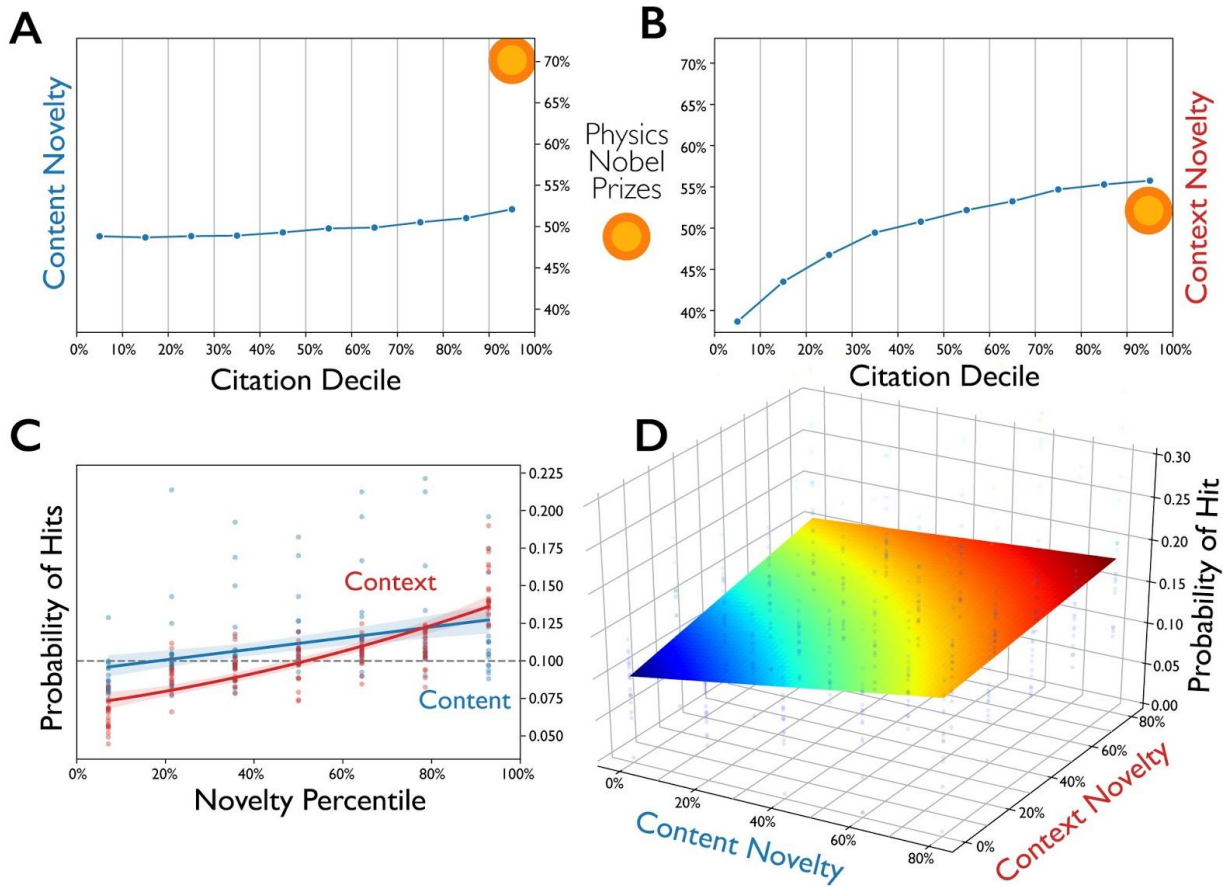

**Fig. S3**. Association between novelty and citation impact or awards for APS papers. Average content (A) and context (B) novelty are plotted for each decile of citations, tracing a monotonic rise; Including averages for Nobel prizes in Physics. Probability of being a hit paper is plotted against content and context novelty separately (C) and jointly (D), manifesting a monotonic increase with novelty.

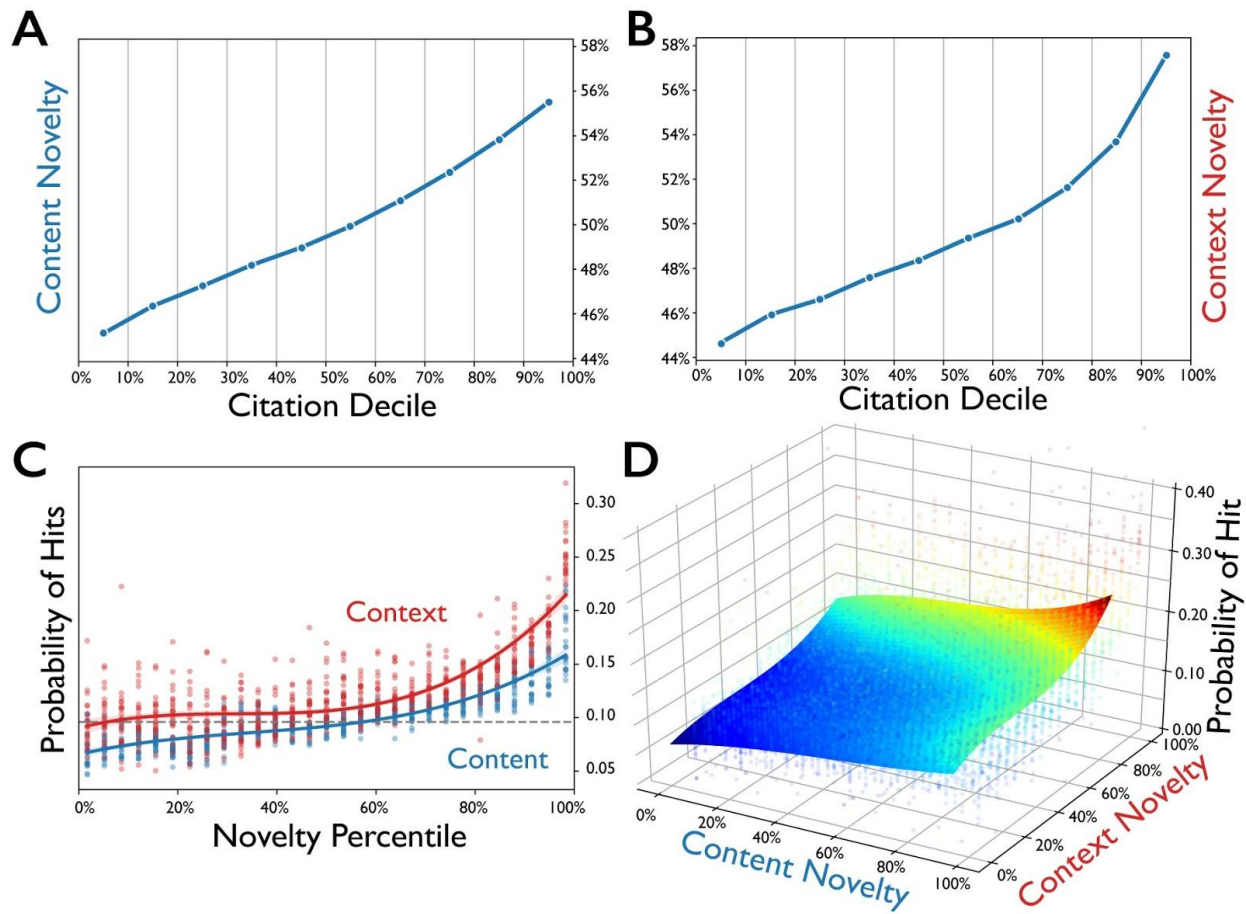

**Fig. S4.** Association between novelty and citation impact for US patents. Average content (A) and context (B) novelty are plotted for each decile of citations, tracing a monotonic rise. Probability of being a hit patent is plotted against content and context novelty separately (C) and jointly (D), manifesting an increase with novelty.

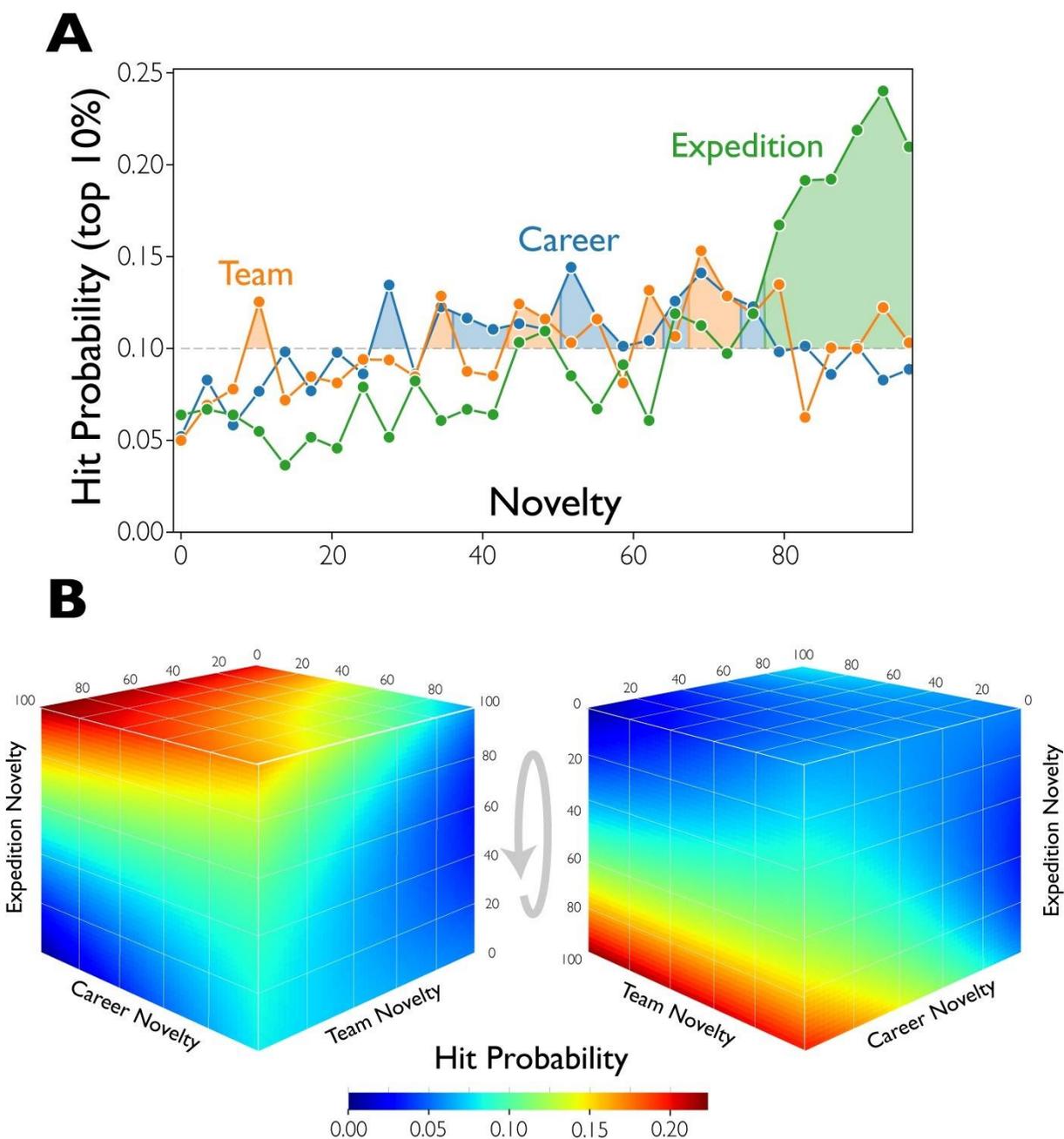

**Fig. S5.** Association between scientists' backgrounds and impact. A: The probability that a hit APS paper was produced by scientists manifesting greater career, team and expedition novelty; with career and team novelty closely correlated and expedition novelty sharply deviating. B: Same as 4B but for APS data. This demonstrates that expedition novelty is the most powerful predictor of outsized success in American physics publications. Team and career novelty play a minor role and are much more uncertain.

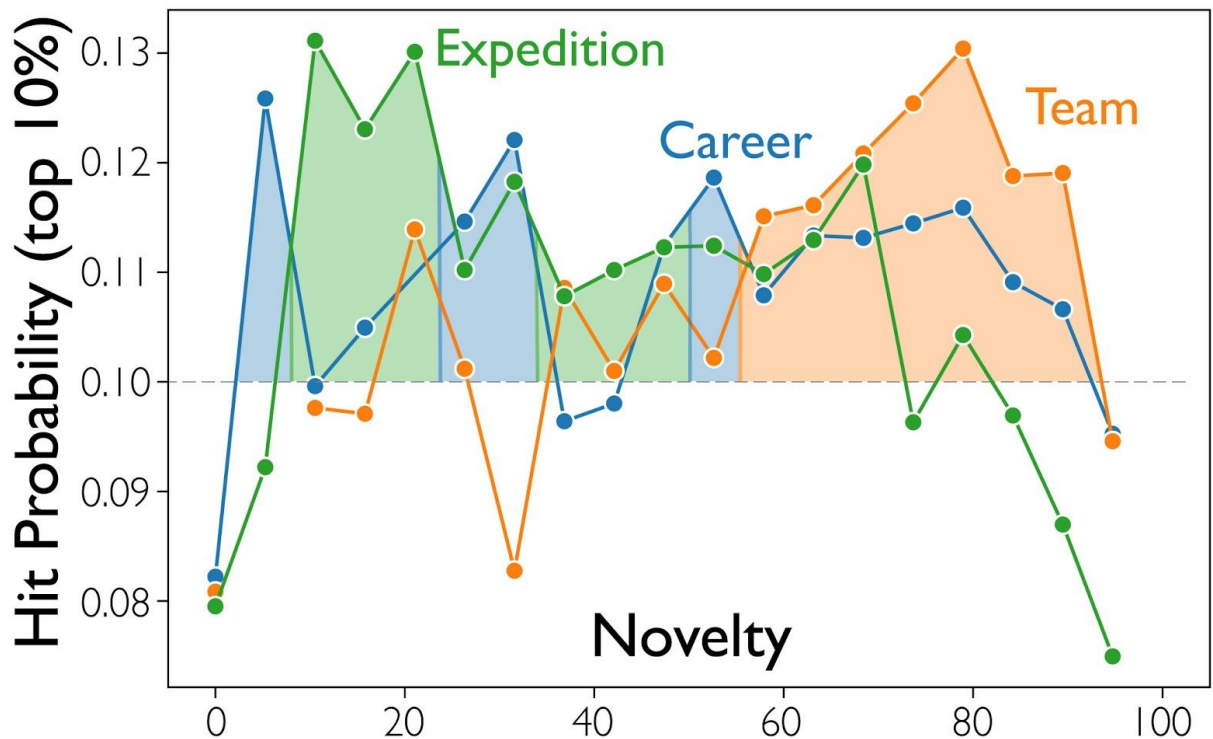

**Fig. S6.** Association between scientists' backgrounds and impact for US patents. The probability that a hit patent was produced by scientists manifesting greater career, team and expedition novelty; suggesting no significant association between hit probability and the background novelties.

# References

1. C. S. Peirce, *Prolegomena to a Science of Reasoning: Phaneroscopy, Semeiotic, Logic* (Peter Lang Edition, 2015).

2. R. K. Merton, E. Barber, The travels and adventures of serenpidity (2004).

3. H. Walpole, Letter from Walpole to Mann, January 28, 1754. *Walpole's Correspondence*. **20**, 407P408 (1754).

4. L. Pasteur, Lecture, University of Lille. *Lille, France. December*. **7**, 1854 (1854).

5. E. Leahey, C. M. Beckman, T. L. Stanko, Prominent but less productive: The impact of interdisciplinarity on scientists' research. *Adm. Sci. Q.* **62**, 105–139 (2017).

6. V. Larivière, Y. Gingras, On the relationship between interdisciplinarity and scientific impact. *J. Am. Soc. Inf. Sci.* **61**, 126–131 (2010).

7. V. Larivière, S. Haustein, K. Börner, Long-distance interdisciplinarity leads to higher scientific impact. *PLoS One*. **10**, e0122565 (2015).

8. H. Youn, D. Strumsky, L. M. A. Bettencourt, J. Lobo, Invention as a combinatorial process: evidence from US patents. *J. R. Soc. Interface*. **12** (2015), doi:10.1098/rsif.2015.0272.

9. W. Brian Arthur, *The Nature of Technology: What It Is and How It Evolves* (Simon and Schuster, 2009).

10. L. Fleming, Recombinant Uncertainty in Technological Search. *Manage. Sci.* **47**, 117–132 (2001).

11. B. Uzzi, S. Mukherjee, M. Stringer, B. Jones, Atypical Combinations and Scientific Impact. *Science*. **342**, 468–472 (2013).

12. L. Fleming, Breakthroughs and the" long tail" of innovation. *MIT Sloan Management Review*. **49**, 69 (2007).

13. T. S. Kuhn, The structure of scientific revolutions, 2nd. *Q. Prog. Rep. United States Air Force Radiat. Lab. Univ. Chic.* (1970).

14. A. R. Benson, D. F. Gleich, J. Leskovec, Higher-order organization of complex networks. *Science*. **353**, 163–166 (2016).

15. J. M. Levine, J. Bascompte, P. B. Adler, S. Allesina, Beyond pairwise mechanisms of species coexistence in complex communities. *Nature*. **546**, 56–64 (2017).

16. J. Grilli, G. Barabás, M. J. Michalska-Smith, S. Allesina, Higher-order interactions stabilize dynamics in competitive network models. *Nature*. **548**, 210–213 (2017).

17. V. Tshitoyan, J. Dagdelen, L. Weston, A. Dunn, Z. Rong, O. Kononova, K. A. Persson, G. Ceder, A. Jain, Unsupervised word embeddings capture latent knowledge from materials science literature. *Nature*. **571**, 95–98 (2019).

18. A. Rzhetsky, J. G. Foster, I. T. Foster, J. A. Evans, Choosing experiments to accelerate collective discovery. *Proc. Natl. Acad. Sci. U. S. A.* **112**, 14569–14574 (2015).

19. T. M. Cover, J. A. Thomas, *Elements of Information Theory* (John Wiley & Sons, 2012).

20. M. Szell, Y. Ma, R. Sinatra, A Nobel opportunity for interdisciplinarity. *Nat. Phys.* **14**, 1075–1078 (2018).

21. F. Sanger, S. Nicklen, A. R. Coulson, DNA sequencing with chain-terminating inhibitors. *Proc. Natl. Acad. Sci. U. S. A.* **74**, 5463–5467 (1977).

22. D. C. Tsui, H. L. Stormer, A. C. Gossard, Two-Dimensional Magnetotransport in the Extreme Quantum Limit. *Phys. Rev. Lett.* **48**, 1559 (1982).

23. A. Gerow, Y. Hu, J. Boyd-Graber, D. M. Blei, J. A. Evans, Measuring discursive influence across scholarship. *Proc. Natl. Acad. Sci. U. S. A.* **115**, 3308–3313 (2018).

24. J. A. Evans, Industry Induces Academic Science to Know Less about More. *Am. J. Sociol.* **116**, 389–452 (2010).

25. T. J. Pinch, W. E. Bijker, The Social Construction of Facts and Artefacts: or How the Sociology of Science and the Sociology of Technology might Benefit Each Other. *Soc. Stud. Sci.* **14**, 399–441 (1984).

26. H. Chesbrough, Open innovation: a new paradigm for understanding industrial innovation. *Open innovation: Researching a new paradigm*. **400**, 0–19 (2006).

27. D. E. Stokes, *Pasteur's Quadrant: Basic Science and Technological Innovation* (Brookings Institution Press, 2011).

28. P. Azoulay, J. Graff-Zivin, B. Uzzi, D. Wang, H. Williams, J. A. Evans, G. Z. Jin, S. F. Lu, B. F. Jones, K. Börner, K. R. Lakhani, K. J. Boudreau, E. C. Guinan, Toward a more scientific science. *Science*. **361**, 1194–1197 (2018).

29. J. Li, Y. Yin, S. Fortunato, D. Wang, A dataset of publication records for Nobel laureates. *Scientific Data*. **6**, 33 (2019).

30. E. M. Airoldi, D. M. Blei, S. E. Fienberg, E. P. Xing, Mixed Membership Stochastic Blockmodels. *J. Mach. Learn. Res.* **9**, 1981–2014 (2008).

31. B. Karrer, M. E. J. Newman, Stochastic blockmodels and community structure in networks. *Phys. Rev. E*. **83**, 016107 (2011).

32. J. G. Foster, A. Rzhetsky, J. A. Evans, Tradition and Innovation in Scientists' Research Strategies. *American Sociological Review*, 80(5), 875–908.

33. I. Torvik, Neil R. Smalheiser, Author name disambiguation in MEDLINE. *ACM Trans. Knowl. Discov. Data* 3, 3, Article 11 (July 2009).

**Acknowledgments:** We are grateful to workshops at MIT Sloan, the Santa Fe Institute and the Computational Social Science program at the University of Chicago for helpful comments. Funding: FS was supported in part by the Data@Carolina initiative; FS and JE were supported by the John Templeton Foundation; JE was supported by the Air Force Office of Scientific Research (FA9550-19-1-0354), National Science Foundation (1829366), and DARPA (HR00111820006). Competing interests: Authors declare no competing interests; Data and materials availability: All data, code, and materials used in the analysis are posted at github.com/knowledgelab/novelty.